\documentclass[envcountsame,runningheads,11pt]{llncs}
\usepackage{a4wide}
\usepackage{xspace}
\usepackage{amsmath}
\usepackage{amssymb}
\usepackage{graphicx}
\newcommand{\bigtimes}{\mathop{\times}}
\newcommand{\IR}{\mathbb{R}}

\newcommand{\IZ}{\mathbb{Z}}

\newcommand{\IN}{\mathbb{N}}
\newcommand{\dom}{\operatorname{dom}}

\newcommand{\diam}{\operatorname{diam}}
\newcommand{\vol}{\operatorname{vol}}
\newcommand{\avrg}{\operatorname{avrg}}

\newcommand{\Prob}{\operatorname{Pr}}
\newcommand{\ball}{B}

\newcommand{\calO}{\mathcal{O}}

\newcommand{\calM}{\mathcal{M}}

\newcommand{\Card}{\operatorname{Card}}
\newcommand{\COMMENTED}[1]{}

\spnewtheorem{observation}[theorem]{Observation}{\bfseries}{\itshape}
\spnewtheorem{fact}[theorem]{Fact}{\bfseries}{\itshape}
\spnewtheorem{myclaim}[theorem]{Claim}{\bfseries}{\itshape}
\spnewtheorem{algorithm}[theorem]{Algorithm}{\bfseries}{\itshape}
\spnewtheorem{scholium}[theorem]{Scholium}{\bfseries}{\itshape}
\spnewtheorem{myexample}[theorem]{Example}{\bfseries}{\itshape}
\spnewtheorem{myremark}[theorem]{Remark}{\bfseries}{\itshape}

\begin{document}
\setcounter{secnumdepth}{3}
\setcounter{tocdepth}{3}
\title{Adaptive Mesh Approach for Predicting Algorithm Behavior\\
with Application to Visibility Culling in Computer Graphics}
\titlerunning{Adaptive Mesh Prediction of Visibility Culling Efficiency}
\author{Matthias Fischer\inst{1} \and Claudius J\"{a}hn\inst{1}\thanks{%
Supported by \textsf{DFG} grants \texttt{Me\,872/12-1} within \textsf{SPP\,1307} and Research Training Group GK-693 of the Paderborn Institute for
Scientific Computation (PaSCo), contact author} \and
Martin Ziegler\inst{2}\thanks{Supported by \texttt{Zi\,1009/2-1}}}
\authorrunning{M.~Fischer, C.~J\"{a}hn, M.~Ziegler}
\institute{Heinz Nixdorf Institute and Department of Computer Science, University of Paderborn, GERMANY
\and Institute for Theoretical Physics, Vienna University of Technology, AUSTRIA}
\date{}
\makeatletter
\renewcommand\maketitle{\newpage
  \refstepcounter{chapter}%
  \stepcounter{section}%
  \setcounter{section}{0}%
  \setcounter{subsection}{0}%
  \setcounter{figure}{0}
  \setcounter{table}{0}
  \setcounter{equation}{0}
  \setcounter{footnote}{0}%
  \begingroup
    \parindent=\z@
    \renewcommand\thefootnote{\@fnsymbol\c@footnote}%
    \if@twocolumn
      \ifnum \col@number=\@ne
        \@maketitle
      \else
        \twocolumn[\@maketitle]%
      \fi
    \else
      \newpage
      \global\@topnum\z@   
      \@maketitle
    \fi
    \thispagestyle{empty}\@thanks
    \def\\{\unskip\ \ignorespaces}\def\inst##1{\unskip{}}%
    \def\thanks##1{\unskip{}}\def\fnmsep{\unskip}%
    \instindent=\hsize
    \advance\instindent by-\headlineindent
    \if@runhead
       \if!\the\titlerunning!\else
         \edef\@title{\the\titlerunning}%
       \fi
       \global\setbox\titrun=\hbox{\small\rm\unboldmath\ignorespaces\@title}%
       \ifdim\wd\titrun>\instindent
          \typeout{Title too long for running head. Please supply}%
          \typeout{a shorter form with \string\titlerunning\space prior to
                   \string\maketitle}%
          \global\setbox\titrun=\hbox{\small\rm
          Title Suppressed Due to Excessive Length}%
       \fi
       \xdef\@title{\copy\titrun}%
    \fi
    \if!\the\tocauthor!\relax
      {\def\and{\noexpand\protect\noexpand\and}%
      \protected@xdef\toc@uthor{\@author}}%
    \else
      \def\\{\noexpand\protect\noexpand\newline}%
      \protected@xdef\scratch{\the\tocauthor}%
      \protected@xdef\toc@uthor{\scratch}%
    \fi
    \if@runhead
       \if!\the\authorrunning!
         \value{@inst}=\value{@auth}%
         \setcounter{@auth}{1}%
       \else
         \edef\@author{\the\authorrunning}%
       \fi
       \global\setbox\authrun=\hbox{\small\unboldmath\@author\unskip}%
       \ifdim\wd\authrun>\instindent
          \typeout{Names of authors too long for running head. Please supply}%
          \typeout{a shorter form with \string\authorrunning\space prior to
                   \string\maketitle}%
          \global\setbox\authrun=\hbox{\small\rm
          Authors Suppressed Due to Excessive Length}%
       \fi
       \xdef\@author{\copy\authrun}%
       \markboth{\@author}{\@title}%
     \fi
  \endgroup
  \setcounter{footnote}{\fnnstart}%
  \clearheadinfo}
\makeatother

\maketitle
\def\thefootnote{\fnsymbol{footnote}}
\addtocounter{footnote}{3}
\begin{abstract}
We propose a concise approximate description, and a method
for efficiently obtaining this description, via adaptive random sampling
of the performance (running
time, memory consumption, or any other profileable numerical
quantity) of a given algorithm on some low-dimensional rectangular
grid of inputs. The formal correctness is proven under reasonable
assumptions on the algorithm under consideration;
and the approach's practical benefit is 
demonstrated by predicting for which observer positions and 
viewing directions an occlusion culling algorithm 
yields a net performance benefit or loss compared to a simple brute force renderer.
\end{abstract}
\begin{minipage}[c]{0.95\textwidth}
\renewcommand{\contentsname}{}
\tableofcontents
\end{minipage}
\pagebreak\addtocounter{page}{-1}%

\section{Introduction} \label{s:intro}

Although it is possible to give bounds for different aspects 
of many algorithms'  runtime behaviors (like for running time or memory consumption) by formal analysis,
the input can heavily influence the actual behavior (e.g. algorithms for real-time rendering
of virtual 3d-scenes). 
To evaluate algorithms for practical applications, in which the input (or the 
characteristics of the input) is known, a more detailed estimation of the algorithm's behavior 
s necessary in order to

\begin{itemize}
	\item select the appropriate algorithm for the given setting (hardware, application and input).
	\item find suitable parameters to adapt the selected algorithm to the setting.
	\item identify bottlenecks and starting points for further improvements of the algorithm.
\end{itemize}
If the number of possible inputs is sufficiently small, it may be possible to evaluate the observed 
property of the algorithm (e.g. the running time) for every input and use this as basis for
the evaluation.
But in most cases the input space is too big (e.g. all possible positions inside a virtual scene), so that 
only a few samples can be evaluated experimentally. 
For a simple uniform sampling approach, it is difficult to capture the structures the input may exhibit.
Another common way in computer graphics is e.g. to select a camera path covering all relevant inputs manually.

But if small changes in the input mostly lead to small changes in the behavior of the algorithm only, we
can apply our adaptive sampling method. Thereby the input space is subdivided into regions, in which
the algorithm behaves similarly. This subdivision can then represent an easy-to-handle model of the algorithm.

In the following, we 
\begin{enumerate}
	\item[i)] present the method that creates this subdivision of the input.
	\item[ii)] prove that, if the function, which describes the behavior of the algorithm is Lipschitz-continuous, it can be approximated by this method.
	\item[iii)] evaluate the method in the domain of real-time 3d rendering, in which we can make use of the distinct local coherence of many rendering algorithms.
\end{enumerate}
%
%
%
%
%
%
The goal of our approach is to preprocess a given algorithm
via blackbox queries at certain inputs,  in order to
quickly and approximately predict its behavior on other inputs.
The data structure constructed during preprocessing is 
a hierarchical subdivision described in Section~\ref{s:subdivision}
and gives a kind of global picture of the algorithm under consideration
with many applications to Algorithm Engineering (Section~\ref{s:tool}).
The success of our approach is both proven formally 
in Section~\ref{s:Lipschitz} (under reasonable analytical assumptions)
and\footnote{Note that we do not claim Occlusion Culling to behave
Lipschitz-continuously, but rather consider ii) and iii) as two 
classes of scenarios that benefit from i)}
demonstrated empirically on the Occlusion Culling problem
in Computer graphics: see
Sections \ref{s:Application}, \ref{s:parameters}, \ref{s:prediction} 
and \ref{s:selection}.


\section{Randomized Adaptive Hierarchical Subdivision}\label{s:subdivision}\label{s:outline}
Consider a $d$-dimensional cuboid 
$C=[a_1,b_1]\times [a_2,b_2]\cdots[a_d,b_d]\subseteq\IR^d$.
We want to approximate an unknown function $f:C\to\IR$,
accessable through blackbox queries for its values $f(\vec x)$
at given arguments $\vec x$,
by a piece-constant function consisting of `few and simple' pieces.
\begin{algorithm} \label{a:Claudius}
Fix dimension $d\in\IN$,
sample size $k$,
so-called \textsf{splitting threshold} $s>0$,
and the unknown $f:C\subseteq\IR^d\to\IR$.
\begin{itemize}
\item Sample $k$ arguments $\vec x_1,\ldots,\vec x_k\in C$
  independently uniformly at random.
\item Query values $y_i:=f(\vec x_i)$, $1\leq i\leq k$.
\item If $|y_i-y_j|\leq s$ holds for all $1\leq i,j\leq k$,\\
  replace $f$ on $C$ by the constant function $g\equiv z:=\avrg(y_1,\ldots,y_k)$.
\item Otherwise cut $C$ into $2^d$ subcuboids of equal size\\
 and recurse to (the restrictions of $f$ to) each of them.
\end{itemize}
\end{algorithm}
The underlying idea is simple: If values $f(\vec x_i)$
and $f(\vec x_j)$ deviate too much, then the cuboid
cannot be accurately described by a constant function
on entire $C$.
\begin{figure}[htb]
\centerline{\includegraphics[width=0.5\textwidth]{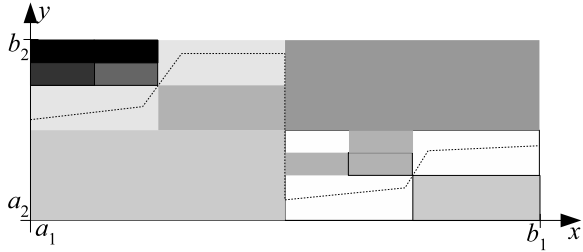}}%
\caption{\label{f:example}Example piecewise constant approximation 
$g$ of a function $f:[a_1,b_1]\times[a_2,b_2]\to\IR$ as output by
Algorithm~\ref{a:Claudius}: Shades of gray indicate the values of $g$.}
\end{figure}
Note that the recursive process of Algorithm~\ref{a:Claudius}
yields a piecewise constant function $g$, in which the locally
constant parts constitute a hierarchical subdivision known
in 2D as \textsf{quadtree} (indicated in Figure~\ref{f:example})
and in 3D as \textsf{octree}.
In particular, this $g$ naturally
comes with a very practical and efficient representation as
a data structure: a hierarchical subdivision.
That is, Algorithm~\ref{a:Claudius} can
be considered as preprocessing $f$ for the following
interpolation:

\begin{algorithm} \label{a:Evaluate}
Fix a $d$-dimensional octree $T$ over $C$
as produced by Algorithm~\ref{a:Claudius}.
\\
Given $\vec x\in C$, iteratively
\begin{itemize}
\item
determine which of the $2^d$ subcuboids $C'$ of $C$
 this input $\vec x$ lies in;
\item proceed to this $C'$ (i.e. to the corresponding
subtree of $T$)
\item unless $C$ is already a leaf of $T$.
\item In the latter case, return the constant
  value $z$ of $g$ on $C$.
\end{itemize}
\end{algorithm}
Because of the exponential term $2^d$, this
approach can 
be efficient only for small values of $d$.
(In Section~\ref{s:Application} we will apply it for $d=2\tfrac{1}{2}$:
two spatial dimensions and a discrete directional one.)
The piecewise constant function $g$,
output by Algorithm~\ref{a:Claudius} 
(and thus also the extrapolated values
produced by Algorithm~\ref{a:Evaluate})
can of course not be expected 
to approximate the given $f$ in general.
But Section~\ref{s:Lipschitz} asserts that it
does so for
Lipschitz-continuous $f$ and sufficiently large
sample sizes.


\subsection{Applications of the Hierarchical Subdivision to Algorithm Engineering} \label{s:tool}
Consider some algorithm $A$ operating on 
(integral or continuous) inputs $\vec x$ from the cuboid $C$.
Let $f(\vec x)$ denote the value at input $\vec x$ of
some profileable quantitative property of $A$.
This may for instance be  running time, 
memory consumption, number of instructions etc.
(Section~\ref{s:Application} will for example
consider the number of occlusion queries.)
Then Algorithm~\ref{a:Claudius} produces a 
hierarchical subdivision of $C$ and piecewise
constant approximation $g$ to $f$ from profiled
values $f(\vec x)$ at `few' 
adaptively sampled inputs $\vec x$;
and Algorithm~\ref{a:Evaluate} uses this data structure
to approximately predict the value $f(\vec x)$
(i.e. the quantitative property of algorithm $A$)
on other inputs $\vec x\in C$.

Now this approach cannot be expected to succeed all the time
and for every $A$. On the other hand many practical
algorithms 
(in particular those operating on continuous,
e.g. floating point data)
do exhibit some sort of continuity---if not
strictly, then in the mollified sense of local averages.
In fact such kind of benignity is analogous to the temporal
and spatial locality hypotheses which both, caching (data)
and instruction prefetch / branch prediction (control) 
techniques successfully do rely on.
And if Algorithm~\ref{a:Claudius} does yield a good
approximation of the behavior of $A$, this
can be employed in various ways:

\textbf{Realistic Rating and Comparison of Algorithms:}
A picture as in Figure~\ref{f:example} summarizes  the behavior
of the underlying algorithm $A$ rather nicely. One 
can easily read off regions of inputs, in which $A$ takes 
long (dark) or little time (bright). A similar one for 
another algorithm $B$ permits to decide for which regions
$A$ may succeed over $B$ and by how much.

\textbf{Average-Case Performance Estimation for Generic Input Distributions:}
Worst-case running times often suffer from few, but practically
rare `bad' inputs. An average case statement 
is restricted to one specific input distribution and may be
useless to another. The hierarchical decomposition
produced by Algorithm~\ref{a:Claudius}
on the other hand allows to estimate average case properties
for many distributions on input space. In fact the
octree needs to be determined only once:
a distribution then amounts to assigning weights to
the sub(sub)cuboids and the induced average case property
to a mere (re-)calculation of their weighted average.
The worst-case can be read off as well.

\textbf{Empirical Algorithm Evaluation on Generic Hardware:}
Algorithm~\ref{a:Claudius} can be applied to the same $A$
in order to determine various profileable quantities $f_i$ separately,
for instance to count how often it performs operation $a_1$,
, $a_2$, etc. . Now suppose that on some 
specific hardware $H$, $a_i$ takes time $t_i$:
Then the total time used by $A$ can be estimated
as $\sum_i f_i\cdot t_i$ 
(for sequential execution; $\max_i f_i\cdot t_i$ for
parallel, and similarly for mixed execution) 
by mere calculation,
i.e. without actually having to execute $A$ on $H$.
%

\textbf{Parameter Optimization:}
Suppose some algorithm $A$ has $d$ input dimensions
and $k$ parameters. A common topic in algorithm engineering
is how to choose these parameters (in dependence of the
specific inputs and hardware) in order to to gain optimum
performance. For example, many tree-based algorithms
behave much better when collapsing all `small' subtrees
of size below a certain threshold into simple arrays.
This threshold value is the parameter to be optimized.
Again, this is a problem which the data structure generated
by Algorithm~\ref{a:Claudius} can help solving.
In Figure~\ref{f:example} for instance, $y$ as a parameter
would be chosen in dependence of input $x$ as indicated by
the dashed line because that yields an overall performance
below the 10\% grey level.

\textbf{Automatic Adaption to Computational Environments}
combines parameter optimization with the performance prediction
from generic to specific hardware.

\subsection{Applications of the Subdivision to the Rendering of Virtual Scenes}\label{s:field}
In this section, we transfer the proposed applications to Algorithm Engineering in the field of real-time 
rendering of 3d-scenes. We give an introduction to the problems and the
algorithms we use for the examples and give an overview over the covered 
applications of the subdivision method in this domain.

%

Our field of interest is the occlusion culling problem in computer graphics:
Interactive display of highly complex scenes like
landscapes with extensive vegetation, large city models, or highly
detailed datasets in scientific visualization are a major challenge
for rendering algorithms.
Although the capabilities of computer graphics hardware have
dramatically increased in the past decades, the handling of scene
complexity is still one of the most fundamental problems in computer
graphics. 
A large number of approaches have been proposed to handle high
scene complexity in interactive applications \cite{RTRbook,LODbook}.
In this paper we focus on occlusion culling:
trying to avoid bothering with rendering parts of the scene
which are not visible to the observer anyway.
A multitude of
algorithms has been suggested and is being employed
for this purpose \cite{vissurvey,visdisssurvey}. This
raises even more questions of whether and under which
circumstances, one algorithm may be superior to the
other.


We apply and evaluate the approach of Section~\ref{s:tool} as a means to solve this problem.
Specifically we consider two fundamentally different rendering algorithms:
the brute-force way of sending all triangles of the scene to the graphics pipeline
and a recent one 
\cite{Bittner}. The latter algorithm uses a feature of modern graphic adapters, which allows counting an object's  number of pixels , which  pass the depth test of the rendering pipeline and are therefore not occluded by a previous object. 
In order to make use of this feature, the virtual scene is organized in a tree that represents a bounding volume hierarchy, in which the axis aligned bounding box of a node encloses the boxes and geometrical objects of all children (here we use an octree). 
The rendering algorithm traverses the nodes of the tree (the nodes are thereby ordered  front-to-back from the observer's position) and before a node is rendered, it's bounding box is tested for if it does contribute at least one pixel to the frame buffer. 
If it's bounding box is not fully occluded, all associated objects are rendered and the traversal continues with the child nodes. If the box is hidden, the corresponding subtree is skipped for this frame since all children can  only lie inside this box, as well and are completely occluded.
As the visibility test itself needs to pass the rendering pipeline, it takes some time for it's result to be available. 
To hide this delay, the algorithm continues with the rendering of the scene and updates the visibility information when the result arrives. 
This can result in the futile rendering of some hidden nodes, but is still faster than locked waiting.

In the evaluation we apply the presented applications (Section  \ref{s:tool}) to these rendering algorithms in the field of computer graphics.
\begin{enumerate}
	\item  We use hierarchical subdivisions to globally compare the algorithms according to their running time and evaluate
	 the occlusion culling efficiency (in Section \ref{s:Application}).
	 \item We use the average values of subdivisions according to running time in order to identify the optimal value for the maximum octree-depth in the example setting (In Section \ref{s:parameters}).
	 \item Section \ref{s:prediction} gives an overview on how the running time can be predicted with subdivisions according to the components of a cost function for generic hardware.
	 \item In Section \ref{s:selection} we present how subdivisions in this area can be extended to include additional information about the viewing direction in order to
	 online select the best rendering algorithm depending on the position and the viewing direction.
\end{enumerate}

%
%
%
%
%
%
%

\subsection{Related Work} \label{s:releated}
There is a vast literature on algorithmic methods
for adaptively and hierarchically approximating an 
unknown function $f$ piecewise by simpler ones.

i) Numerical treatment of a function $f$
on some smooth manifold $\calM$ 
(e.g. the solution to a partial differential equation
using the \emph{Finite Elements Method} \textsf{FEM})
generally proceeds by 
first triangulating $\calM$ and replacing $f$
on each such triangle by a linear function.
For reasons of accuracy, the thus considered mesh on $\calM$ 
is usually desired to be finer on regions of high curvature
(corresponding to large variations of $f$)
and coarse on `flat' parts of $\calM$;
cf. e.g. \cite{AdaptiveMeshRefinement}.

ii) Also well-known are various methods of approximately `learning'
an unknown function $f$ by querying its values $f(\vec x)$
on appropriately chosen arguments $\vec x_i$. This is in fact
the essence of information-based complexity \cite{Traub},
e.g. for numerically integrating $f$.
Again, the evaluation points are preferably chosen
adaptively: more densely if, and where, $f$ 
exhibits strong variations.

A synthesis of i) and ii)
concerning Lipschitz-continuous functions
(cf. Section~\ref{s:Lipschitz} below),
the work \cite{Demaine} is concerned with 1D integration;
and \cite{Cooper,Beliakov} focus on
deterministic uniform approximation of $f$
by piecewise linear functions.
We point out that i+ii) can also be considered as 
lossy function compression problems: replacing a 
(possibly complicated)
$f$ by some simple $g$ resembling $f$. 
`Simple', here means, piecewise constant or linear;
for other classes of simple functions (like, e.g.,
sines and cosines) one arrives at wavelets and
Fourier compression with famous applications such as
\texttt{mp3} and \texttt{jpeg}.

The main idea of the present work is to apply these
methods to a seemingly unrelated but notorious problem
in Algorithm Engineering in general and especially in computer graphics:

iii) Predicting the behavior of an algorithm, e.g., runtime.
This has been a major topic particularly in parallel 
and distributed computing---compare e.g.
\cite{MafiDipl,OlafPubl,Blanco}---and to computer graphics \cite{adaptive,Wonka}.

iv) Evaluating the efficiency (runtime, frames) of a rendering
algorithm. Typically, the target function is measured along a chosen
camera path \cite{adaptive} or with an increasing scene complexity
(number of polygons) \cite{spatial}
or for some fixed chosen viewing points \cite{surfels}.

v) To maintain an adaptive rendering algorithm. This has been applied 
to real-time rendering systems to adaptively adjust image quality 
in oder to maintain a uniform, user-specified target frame rate \cite{adaptive}.

%

\section{Asymptotic Analysis and Correctness of the Approach} \label{s:Lipschitz}
As already mentioned, Algorithms~\ref{a:Claudius} and \ref{a:Evaluate}
cannot be expected to succeed on an arbitrary
unknown function. We now prove that they do 
approximate such $f:C\to\IR$ with high
probability, if $f$ is Lipschitz-continuous,
provided that the sample size is large enough.
More precisely 
Theorem~\ref{t:Lipschitz} asserts successful approximation
up to given absolute error
\begin{itemize}\itemsep0pt%
\item uniformly 
  for sample size roughly proportional to 
  the volume of $C$
\item in the least squares-sense
  for sample size roughly proportional to
  the diameter of $C$.
\end{itemize}
Both are shown asymptotically best possible.

\subsection{Reminder and Properties of Lipschitz-continuous Functions}
For $c>0$, a classical notion in
calculus calls a function $f:\dom(f)\subseteq\IR^d\to\IR$
\emph{$c$-Lipschitz}
if  $|f(\vec x)-f(\vec y)|\leq c\cdot\|\vec x-\vec y\|$
holds for all $\vec x,\vec y\in\dom(f)$.
For instance 
any differentiable $f$ with derivative bound
$\|f'(\vec x)\|'\leq c$
is $c$-Lipschitz. Here $(\|\,\cdot\,\|,\|\,\cdot\,\|')$ 
denotes a dual pair of norms, which is satisfying 
the H\"{o}lder Inequality
$|\langle\vec x,\vec y\rangle|\leq\|\vec x\|\cdot\|\vec y\|'$
for all $\vec x,\vec y\in\IR^d$; e.g. Euclidean norms, or
$\|\vec x\|=(\sum_i |x_i|^p)^{1/p}$ and
$\|\vec y\|'=(\sum_i |y_i|^q)^{1/q}$ 
with $1=1/p+1/q$.

We also remark that any continuous function on $C$
is $c$-Lipschitz for some (but possibly very large) $c$,
because $C$ is compact.

\begin{figure}[htb]
\hfill\includegraphics[width=0.29\textwidth]{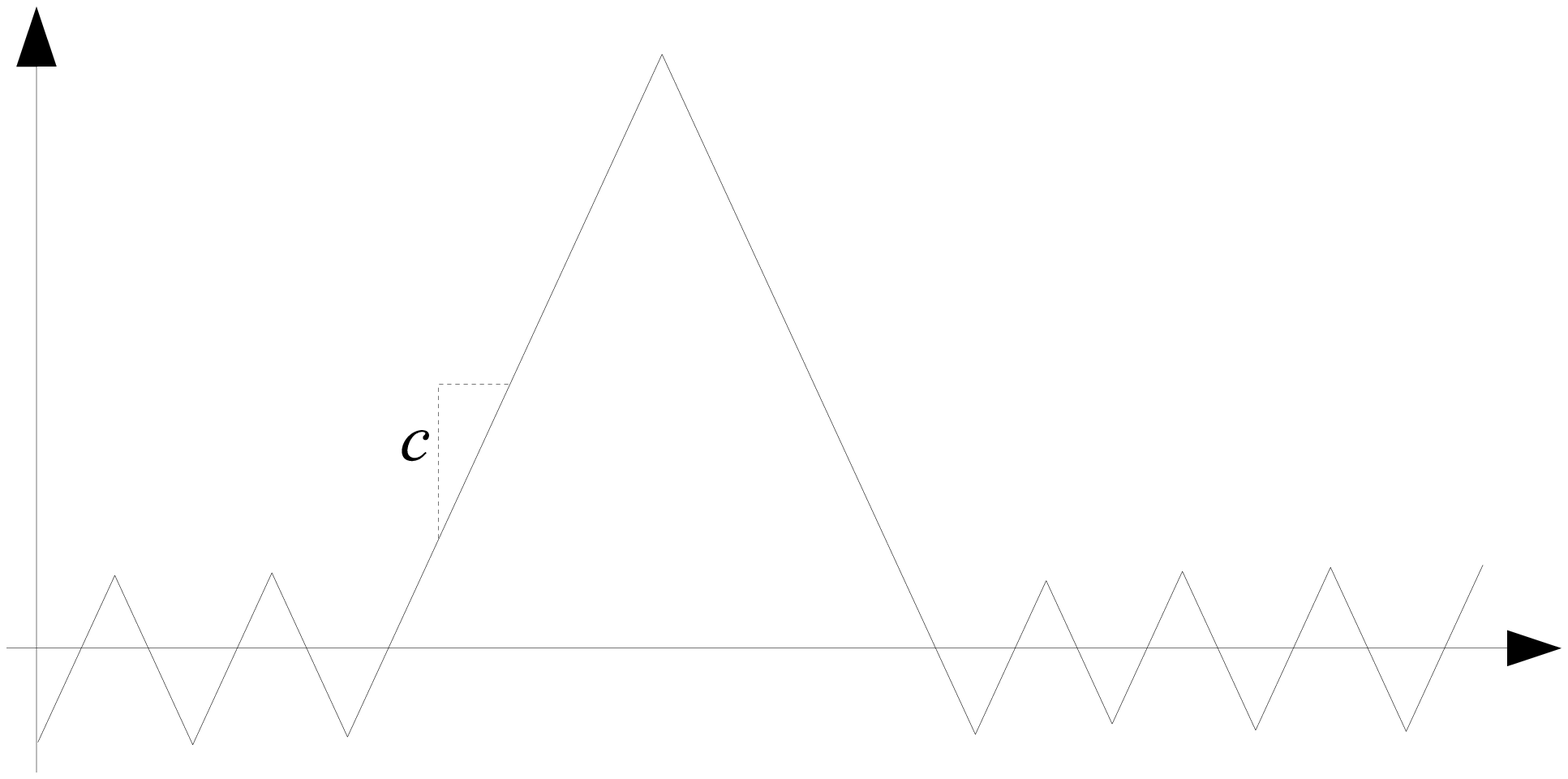}\hfill
\includegraphics[width=0.29\textwidth]{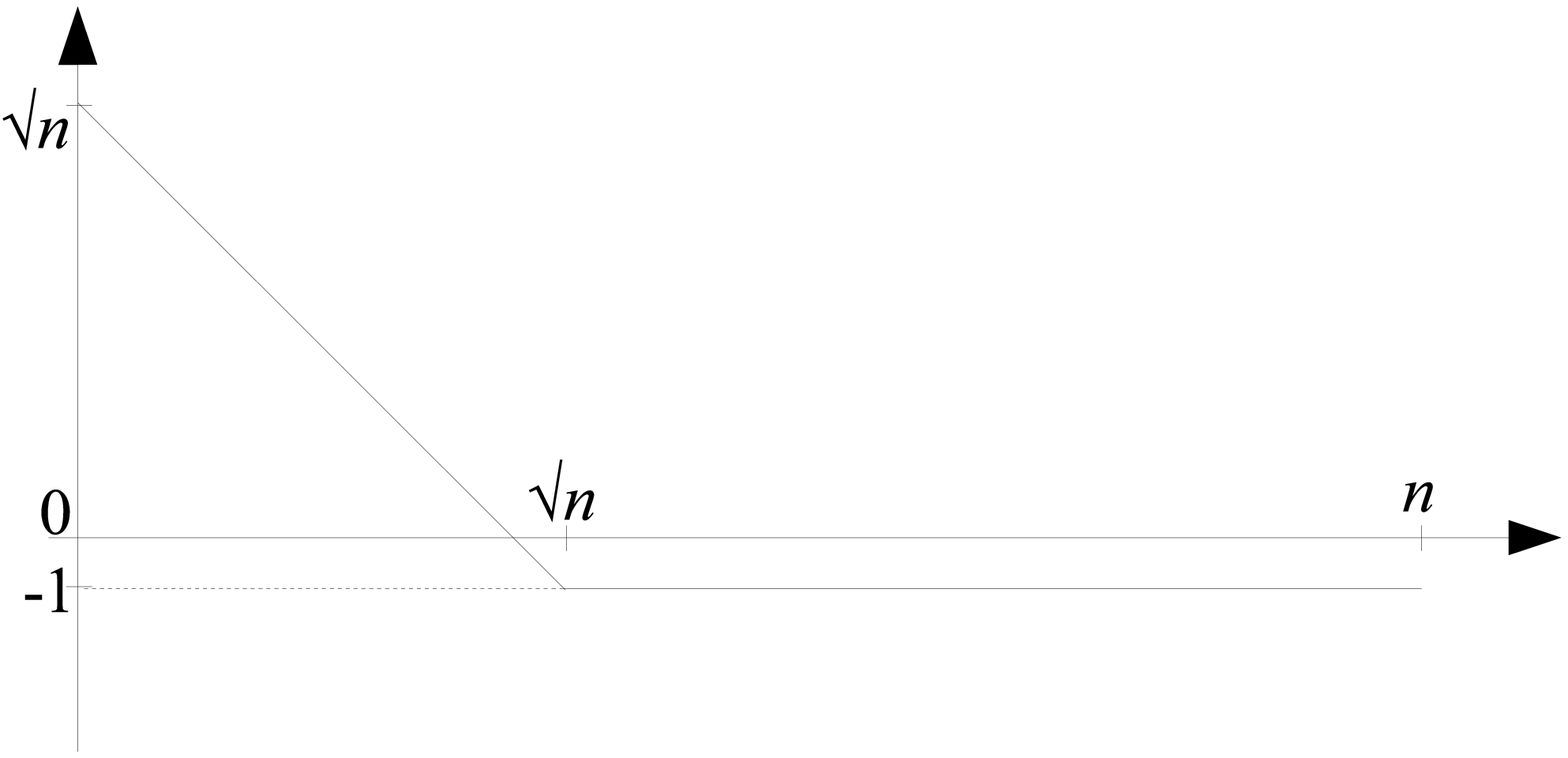}\hfill~
\caption{\label{f:badlipsch}a) A sample of size 
$\Omega(\text{cube volume}/\text{Lipschitz constant})$
is generally necessary in order to get 
uniform approximation of an unknown $f$.
b) 1-Lipschitz function on $[0,n]$ with 
$\int|f|=\Theta(n)$ and $\int f^2=\Theta(n^{3/2})$}
\end{figure}

\begin{lemma} \label{l:Lipschitz}
Let $\mu$ denote a probability measure on 
$C=[a_1,b_1]\times [a_2,b_2]\cdots[a_d,b_d]\subseteq\IR^d$,
i.e. with $1=\mu(C)=\int_C 1\,d\mu$. Moreover
write $\diam(C):=\sup\{\|\vec x-\vec y\|:\vec x,\vec y\in C\}$
and $\ball(\vec x,r):=\{\vec y\in C: \|\vec x-\vec y\|\leq r\}$.
Finally consider a $c$-Lipschitz function $f:C\to\IR$.
\begin{enumerate}
\item[a)]
For $s>0$ and $\vec y\in C$, it holds:
$\displaystyle \big\{\vec x\in C:|f(\vec x)-f(\vec y)|\leq s\big\}
\supseteq\ball(\vec y,s/c)$.
\item[b)]
Let $\lambda:=\min_{\vec y\in C}\mu\big(\ball(\vec y,s/c)\big)$ and suppose
$\sup_C f-\inf_C f>4s$. Then $k:=1/\lambda$ points $\vec x$,
sampled from $C$ according to the distribution $\mu$, contain with
constant probability some $\vec x_i,\vec x_j$ such that 
$|f(\vec x_i)-f(\vec x_j)|>s$.
\item[c)]
For any $\vec y\in C$ and $s>0$
and measurable $C'\subseteq C$, it holds
\[
\int\nolimits_{C'} |f(\vec x)-f(\vec y)|\:d\mu(\vec x)
\quad\leq\quad
s\;\;+\;\; c\cdot\diam(C)\cdot\mu\{\vec x\in C':|f(\vec x)-f(\vec y)|> s\}  \enspace . 
\]
\item[d)]
Let $\mu$ either be the normalized Lebesgues measure on $C$
or the normalized integer counting measure on $C$,
i.e. $\mu(C')=\Card(C'\cap\IZ^d)/\Card(C\cap\IZ^d)$.
Then it holds
\[ \int\nolimits_C f(\vec x)^2\,d\mu(\vec x) \quad\leq\quad
\calO(\sqrt{\diam C})\cdot
\Big(\int\nolimits_C |f(\vec x)|\,d\mu(\vec x)\Big)^2 \quad+\quad \calO(c^2) \enspace . \]
\end{enumerate}
\end{lemma}
Concerning d) remember that, without Lipschitz-condition, $\int_0^1 f(x)^2\,dx$
in general cannot be bounded in terms of $\int_0^1 |f(x)|\,dx$: 
consider $x\mapsto 1/\sqrt{x}$. Also, $\sqrt{\diam C}$ is
asymptotically best possible in 1D, since the 1-Lipschitz function
on $[0,n]$
depicted in Figure~\ref{f:badlipsch}b) has, for the
normalized Lebesgues measure $d\mu=dx/n$, 
$\int|f|\,d\mu=\theta(1)$ and $\int|f|^2\,d\mu=\Theta(n^{1/2})$.
Similarly, the sample size of Lemma~\ref{l:Lipschitz}b) is asymptotically
best possible for the function depicted in Figure~\ref{f:badlipsch}a).

\begin{proof}[Lemma~\ref{l:Lipschitz}]
~a) Since $f$ is $c$-Lipschitz, $\|\vec x-\vec y\|\leq s/c$
implies $|f(\vec x)-f(\vec y)|\leq s$.
\begin{enumerate}\itemsep0pt%
\item[b)]
Let $z:=(\sup f+\inf f)/2$, $\vec y_+,\vec y_-\in C$ with
$f(\vec y_+)=\sup f$ and $f(\vec y_-)=\inf f$,
exploiting that $f$ is continuous on compact $C$.
Then $C_+:=\{\vec x\in C:f(\vec x)>z+s\}
\supseteq\{\vec x:|f(\vec x)-f(\vec y_+)|\leq s\}$
and $C_-:=\{\vec x:f(\vec x)<z-s\}$ both have
$\mu(C_+),\mu(C_-)\geq\lambda$ by a).
Hence $1/\lambda$ points $\vec x$ sampled from $C$
have probability $1-(1-\lambda)^{1/\lambda}\geq1-1/e$
of hitting some $\vec x_i\in C_+$;
similarly for hitting some $\vec x_j\in C_-$:
and since $C_+\cap C_-=\emptyset$ are independent,
a constant probability $\geq(1-1/e)^2$ for both.
These satisfy $|f(\vec x_i)-f(\vec x_j)|>s$.
\item[c)] Observe that probability measure satisfies
$\mu\{\vec x\in C':|f(\vec x)-f(\vec y)|\leq s\}\leq\mu(C)=1$
and
\[
\int\limits_{C'} |f(\vec x)-f(\vec y)|\:d\mu(\vec x)
\quad\leq\quad
\int\limits_{\makebox[4em][l]{$\scriptscriptstyle
\vec x\in C':|f(\vec x)-f(\vec y)|\leq s$}}
s\,d\mu(\vec x)
\quad+\quad
\int\limits_{\makebox[4em][l]{$\scriptscriptstyle
\vec x\in C':|f(\vec x)-f(\vec y)|> s$}}
c\cdot\|\vec x-\vec y\|\,d\mu(\vec x)
\]
\item[d)]
W.l.o.g. suppose $f:[0,N)\to\IR$ is 
nonnegative and $1$-Lipschitz; otherwise consider $|f|/c$.
Consider the real sequence $\vec y$
defined by $y_0:=\max\{1,f(0)\}$, $y_1:=\max\{1,f(y_0)\}$,
$y_2:=\max\{1,f(y_0+y_1)\}$,
$y_{i+1}:=\max\{1,f(y_0+y_1+\cdots+y_i)\}$.
Since $y_i\geq1$, it holds $\sum_{i\leq n} y_i\geq N$ 
for some $n\leq N$; truncate $y_n$ such that 
$\sum_{i<n}<\sum_{i\leq n} y_i=N$.
Moreover let $I\subseteq\{1,\ldots,n\}$
denote the set of those $i$ with $y_i>1$
and in particular $y_i=f(\sum_{j<i} y_j)$.
Because of the Lipschitz condition,
it holds 
\[ y_i-x\leq f(\sum_{j<i} y_j+x)\overset{(*)}{\leq} y_i+x
\quad\text{ for all $0\leq x<y_i$ and all $i\in I$; 
\quad (*) even for all $i$.} \]
Thus,
$\int_0^{y_i} f(\sum_{j<i} y_j+x)\,dx\geq y_i^2/2$ for $i\in I$
and $\int_0^{y_i} f(\sum_{j<i} y_j+x)^2\,dx\leq y_i^3\cdot7/3$
for all $i$; in particular $\leq 7/3$ for $i\not\in I$.
Therefore,
\[ 1/N\cdot\int_0^N|f|\,dx 
\;\;=\;\;1/N\cdot\sum\limits_i\int_{\sum_{j<i} y_j}^{\sum_{j\leq i} y_j} |f|\,dx
\;\;\geq\;\;\tfrac{1}{2}\cdot\sum\limits_{i\in I} y_i^2/\sum\limits_{i\in I} y_i
\;\;=\;\;\tfrac{1}{2}\|\vec y\|_2^2/\|\vec y\|_1 
\] 
with the abbreviation $\vec y:=(y_i)_{_{i\in I}}$. Similarly,
\begin{eqnarray*}
1/N\cdot\int_0^N|f|^2\,dx 
&\leq&1/N\cdot\sum_{i\in I} \int_{\sum_{j<i} y_j}^{\sum_{j\leq i} y_j} |f|^2\,dx
\;+\;1/N\cdot\sum_{i\not\in I} 7/3 \\
&\leq& \tfrac{7}{3}\|\vec y\|_3^3/\|\vec y\|_1 \;+\;7/3
\;\overset{(**)}{\;\leq\;}\; \tfrac{7}{3}\sqrt{N}\cdot\|\vec y\|_2^4/\|\vec y\|_1^2\;+\;7/3
\end{eqnarray*}
where at (**) we have employed Jensen's Inequality
$\|\vec y\|_3\leq\|\vec y\|_2$ as well as the bound
$\|(y_1,\ldots,y_k)\|_1\leq\sqrt{k}\|(y_1,\ldots,y_k)\|_2$.
\qed\end{enumerate}\end{proof}
\subsection{Sample Sizes for Uniform and for Residual Sum-of-Squares Approximation}
First note that Algorithm~\ref{a:Claudius} terminates
for $c$-Lipschitz continuous functions $f:C\to\IR$:
If $c\cdot\diam(C)\leq s$ holds, then line 3 strikes;
and this happens latest at a recursion depth
of order $\log\big(c\cdot\diam(C)/s)$.

\begin{theorem} \label{t:Lipschitz}
Let $\mu$ denote either the normalized Lebesgues measure
or the normalized integer counting measure on $C$.
\begin{enumerate}
\item[i)] 
Consider $k':=\vol(C)\cdot c^d/s^d$
with $\vol(\bigtimes_i [a_i,b_i]):=\prod_{i=1}^d (b_i-a_i)$.
Then Algorithm~\ref{a:Claudius} 
with $k:=k'\cdot\log^2(k')$
produces $g$ such that $\|f-g\|_{\infty}\leq4s$ 
holds with high probability;
\item[ii)] Suppose $s>c$ and 
consider $k':=\sqrt{\diam C}+c\cdot\diam(C)/s$.
Then Algorithm~\ref{a:Claudius} 
with $k:=k'\cdot\log^2(k')$
produces $g$ such that $\|f-g\|_{2}\leq4s$ 
holds with high probability.
\end{enumerate}
\end{theorem}
We remark that the uniform approximation in i)
is based on samples of size $k$, essentially
linear in $\vol(C)$;
whereas the least squares approximation in ii)
takes samples of size proportional to
the diameter---which is asymptotically
smaller in dimensions $d\geq2$.
Note that a sample size of order
$\diam^{1-\epsilon}/s$ is necessary
in the worst case:

\begin{myexample} \label{x:Lipschitz}
Fix $\epsilon>0$ and generalize the 
function in Figure~\ref{f:badlipsch}b) to 
1-Lipschitz
$f_\epsilon:[0,n]\to[0,n]$ defined by
\[ f(x):=n^{1-2\epsilon}-x \quad\text{for }
x\leq m:=n^{1-2\epsilon}+a\cdot n^{1-4\epsilon}, \qquad
f(x):=-a\cdot n^{1-4\epsilon}
\quad\text{for }x\geq m \enspace . \]
Then $\int_0^{n^{1-2\epsilon}} f(x)\,dx=n^{2-4\epsilon}/2$,
$\int_{n^{1-2\epsilon}}^m f(x)\,dx=-a^2\cdot n^{2-8\epsilon}/2$,
and $\int_m^1 f(x)\,dx=-(n-m)\cdot a\cdot n^{1-4\epsilon}$;
hence the mean is $\int_0^n f(x)=0$ 
for some appropriate $a=\tfrac{1}{2}+o(1)$.
Moreover $\int_0^{n^{1-2\epsilon}} |f(x)|^2\,dx
=n^{3-6\epsilon}/3$ and
$\int_{n^{1-2\epsilon}}^n |f(x)|^2\,dx
=\Theta(n^{3-8\epsilon})$, hence the variance
is $\sigma=\sqrt{1/n\cdot\int_0^n |f|^2\,dx}
=\Theta(n^{1-3\epsilon})$.
Now consider $s:=\sigma/2$ and observe that,
in order to
approximate $f$ up to error $s$,
any algorithm has to distinguish it
from some other function $f':[0,n]\to[\pm s/2]$
and therefore needs to detect
$x_i,x_j$ with $f(x_i)-f(x_j)\geq s$.
Since $f(x)\in[-\Theta(n^{1-4\epsilon}),0]\subseteq[-s/2,+s/2]$
for $x\geq n^{1-2\epsilon}$ and $n$ sufficiently large, 
such an algorithm must in particular (yet does not suffice to)
find at least one $x_i\leq n^{1-2\epsilon}$:
which for one single sample happens with probability
$n^{1-2\epsilon}/n
=\Theta(s/n^{1-\epsilon})$
where $n=\diam[0,n]$.
\qed\end{myexample}

\begin{proof}[Theorem~\ref{t:Lipschitz}]
\begin{enumerate}
\item[i)]
First observe that in case $\sup_C f-\inf_C f\leq4s$,
any sampled $\vec x\in C$ will yield $g:\equiv f(\vec x)$
with $\|f-g\|_\infty\leq4s$; whereas in case
$\sup_C f-\inf_C f>4s$, Algorithm~\ref{a:Claudius} has
a high probability of cutting $C$ into smaller parts:
this follows from Lemma~\ref{l:Lipschitz}b) 
by probability amplification
due to the logarithmic oversampling. 
Since the algorithm has logarithmic recursion depth
with sub-subcuboids of exponentially fast decreasing size 
before arriving at constant volume,
the $\log^2$-factor maintains the high success probability
throughout.
\item[ii)]
The \textsf{mean} $z:=\int_{C} f(\vec x)\,d\mu(\vec x)$
is well-known 
to minimize $\int_C|f(\vec x)-z|^2\,d\mu(\vec x)=\|f-z\|_2^2$
(or, equivalently, $\|f-z\|_2$).
Observe that $C_+:=\{\vec x\in C:f(\vec x)\geq z\}$ 
and $C_-:=\{\vec x\in C:f(\vec x)<z\}$ satisfy
$1=\mu(C_-)+\mu(C_+)$ and
\begin{eqnarray*}
0\;=\;\int_{C} \big(f(\vec x)-z\big)\,d\mu(\vec x) 
&\;\;=\;&
\int_{C_+} |f(\vec x)-z|\,d\mu(\vec x)
\;\;-\;\;
\int_{C_-} |f(\vec x)-z|\,d\mu(\vec x) \enspace , \\
\int_{C} |f(\vec x)-z|\,d\mu(\vec x) 
&\;\;=\;&
\int_{C_+} |f(\vec x)-z|\,d\mu(\vec x)
\;\;+\;\;
\int_{C_-} |f(\vec x)-z|\,d\mu(\vec x)
\enspace ; \end{eqnarray*}
hence $\int_{C_\pm} |f(\vec x)-z|\,d\mu(\vec x)
=\tfrac{1}{2}\int_C |f(\vec x)-z|\,d\mu(\vec x)$.
W.l.o.g. $\mu(C_+)\geq1/2$.
\\
Now first suppose $s>\tfrac{1}{4}\|f-z\|_2/\sqrt[4]{\diam C}$.
Recall \textsf{Bernstein's Inequality}
\[
\Prob\Big[\;\big|\sum\nolimits_{i=1}^k X_i/k - z\big|\:>\:s\;\Big]
\quad\leq\quad 2\cdot\exp\Big(-\frac{ks^2}{2\sigma^2+(b-a)\cdot s}\Big) 
\]
for $k$ independent random variables $X_i\in[a,b]$
with mean $z$ and variance $\sigma=\|f-z\|_2$.
Here $X_i:=f(\vec x_i)\in[z\pm c\cdot\diam C]$
because of the Lipschitz condition.
Hence $k\geq32\cdot\sqrt{\diam C}+c\cdot\diam(C)/s$
samples suffice for the sample average 
in line 4 of Algorithm~\ref{a:Claudius}
to be closer than $s$ to the true mean $z$
with constant probability.
\\
This time suppose 
$s>\tfrac{1}{4}\|f-z\|_2/\sqrt[4]{\diam C}$.
Then, by Lemma~\ref{l:Lipschitz}c),
\begin{multline*}
\mu(\{\vec x\in C:f(\vec x)<z-s\})
\quad=\quad\mu(\{\vec x\in C_-:|f(\vec x)-z|>s\}) \quad\geq \\
\quad\geq\quad
\Big(\int\nolimits_{C_-} |f(\vec x)-z|\,d\vec x\;-\;s\Big)
/(c\cdot\diam C) 
\quad=\quad \frac{\|f-z\|_1/2 - s }{c\cdot\diam C}
\end{multline*}
where,
according to Lemma~\ref{l:Lipschitz}d) and in the big-Oh sense,
\[ \|f-z\|_1 
\;\geq\;\sqrt{\|f-z\|^2_2-c^2}/\sqrt[4]{\diam C}
\;\geq\; (\|f-z\|_2-c)/\sqrt[4]{\diam C} 
\;\geq\; 4s - c/\sqrt[4]{\diam C} 
\enspace , \] 
hence $\mu(\{\vec x\in C:f(\vec x)<z-s\})
\geq(3s-c)/(c\cdot\diam C)$.
So a sample of size $k\geq (c\cdot\diam C)/(3s-c)$
has constant probability for 
Algorithm~\ref{a:Claudius} to 
find some $\vec x_i\in C$ 
with $f(\vec x_i)<z-s$ and some
$\vec x_j\in C_+$, i.e. to
proceed to its fourth line.
\qed\end{enumerate}\end{proof}

\section{Empirical Evaluation and Comparison of Occlusion Culling Algorithms} \label{s:Application}

Our approach based on hierarchical subdivision introduces a powerful alternative or addition to the
standard way of evaluating a target function (e.g. the running time) of a rendering algorithm 
along some (hopefully carefully chosen) camera path or at certain observer positions. 

In order to determine 
whether or not this algorithm is appropriate for a specific application,
it is necessary to evaluate the algorithm with respect to  different requirements. 
One such requirement for an occlusion culling algorithm could be for example 
that it actually identifies and rejects a sufficiently large part 
of the hidden objects during the rendering process. 
More relevantly: Does the algorithm increase the overall frame rate 
of the scene on the target system compared to a simpler algorithm 
without occlusion culling, 
taking into account the overhead introduced by the occlusion tests.

\begin{figure}
\begin{minipage}[t]{0.32\textwidth}
      \centering
		   \includegraphics[width=0.80\textwidth]{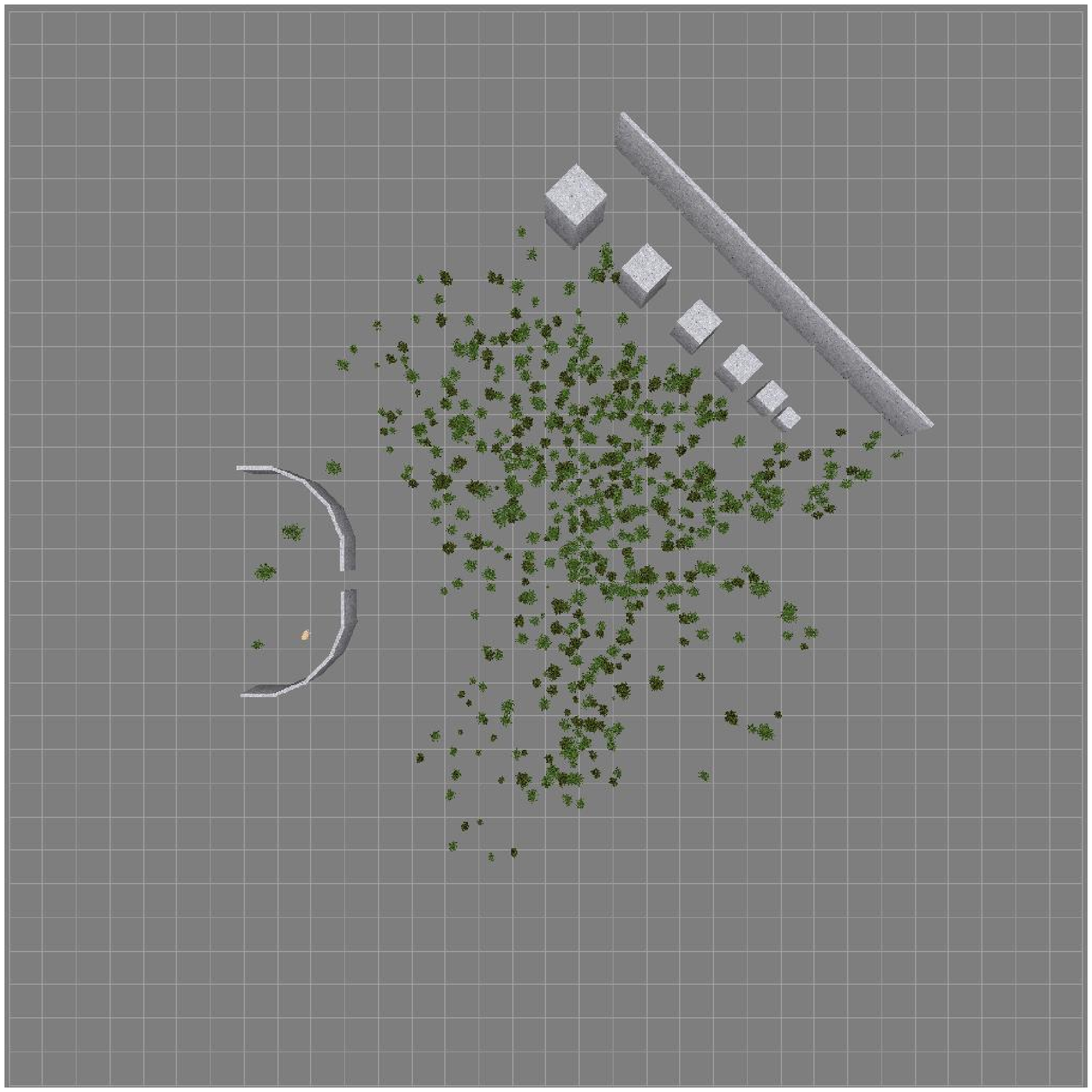}
       \caption{Example scene consisting of 625 objects (6.2M polygons)}
       \label{fig:scene}
\end{minipage}%
\hspace{2mm}
\begin{minipage}[t]{0.32\textwidth}
      \centering
      \includegraphics[width=0.80\textwidth]{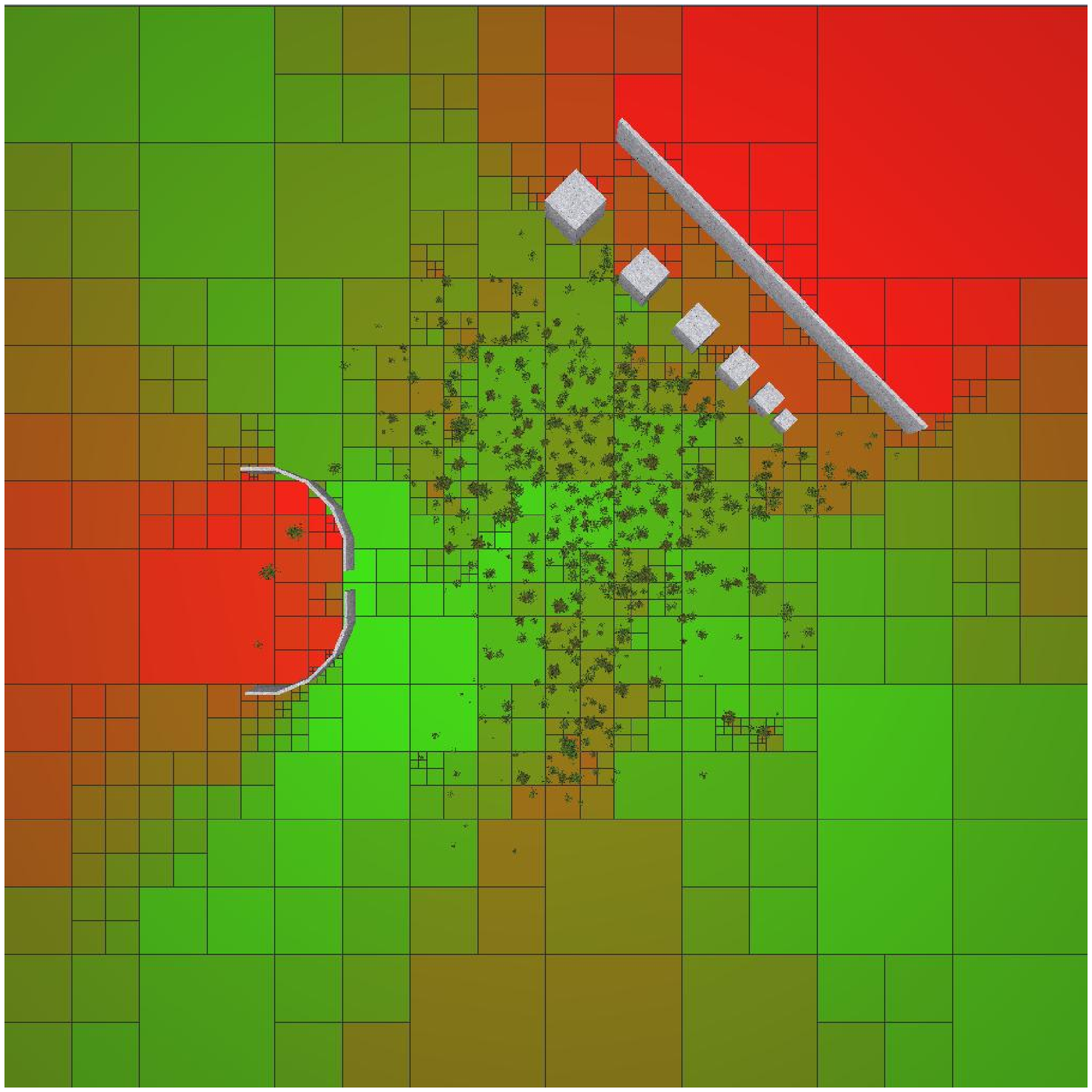}
      \caption{Subdivision according to the number of visible objects (resolution $256\times256\times1$, splitting threshold 40)}
      \label{fig:part_vis}
\end{minipage}%
\hspace{2mm}
\begin{minipage}[t]{0.32\textwidth}
      \centering
      \includegraphics[width=0.80\textwidth]{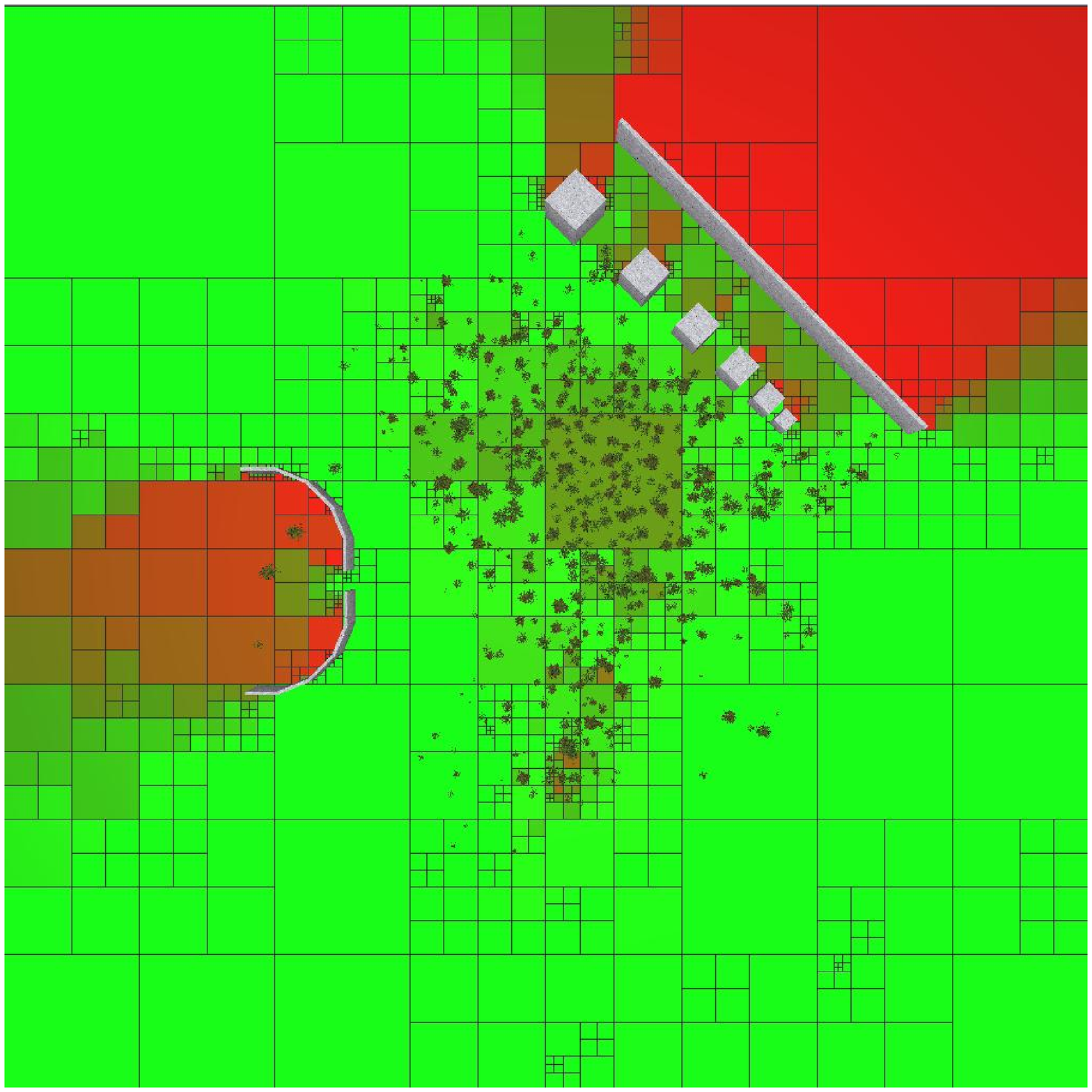}
      \caption{Subdivision according to the number of as visible classified objects (resolution $256\times256\times1$, splitting threshold 80)}
      \label{fig:part_vis_occ}
\end{minipage}
\end{figure}			
Figure \ref{fig:scene} shows the scene we use for the following examples. 
It consists of 625 Objects (of 6.2 million triangles altogether), mostly trees (which individually produce only little occlusion) and some opaque walls. 
To inspect the functionality of our chosen occlusion culling algorithm in this setting, 
we construct the hierarchical subdivision 
with respect to the function ``number of visible objects''. 
Visibility is determined by projecting (rendering) the scene onto the sides 
of a cube around the current position for which the (visibility-)function 
is evaluated. An object is considered visible if it contributes at least one pixel to the rendered image of one of the six sides of the cube.

In our implementation of the sampling approach, we restrict the sample space to a discrete set of points in 3d-space arranged on a grid, so that the maximum size of the subdivision is bounded by the resolution of the grid (the smallest cell of a subdivision is one cell of the grid). This resolution can be chosen as one parameter to adapt the resulting subdivision to the needs of the intended application. The sampling size is set proportional to the diameter of the current area in grid points (times 0.5).

The created subdivision with a grid-resolution  of $256\times256\times1$ (in this example we inspect only one layer) is shown in Figure \ref{fig:part_vis}; the areas where only few objects are visible are colored red and the areas where almost all objects are visible are colored green. 
The three dimensional visualization of the data gives a very intuitive impression of the actual distribution of the scene's visibility function. 
To test how many of the hidden objects of this scene our occlusion culling algorithm identifies, we create an additional subdivision according to the number of objects the algorithm classifies as visible. The value of the function is measured by executing the rendering algorithm and counting the objects that pass the occlusion test on at least one side of the cube.

Just by comparing the resulting image (Figure \ref{fig:part_vis_occ}) to the previous subdivision according to the visibility you can see that in areas where many objects are occluded, the number of rendered objects is indeed smaller; but the number of rendered objects is in general larger than the number of visible objects.
This can be further analyzed by creating an additional subdivision of the difference between the visible and rendered objects (which can easily be calculated from the existing subdivisions). This \textit{difference subdivision} indicates the number of unnecessarily rendered objects.

\begin{figure}
\begin{minipage}[t]{0.5\textwidth}
      \centering
		   \includegraphics[width=0.80\textwidth]{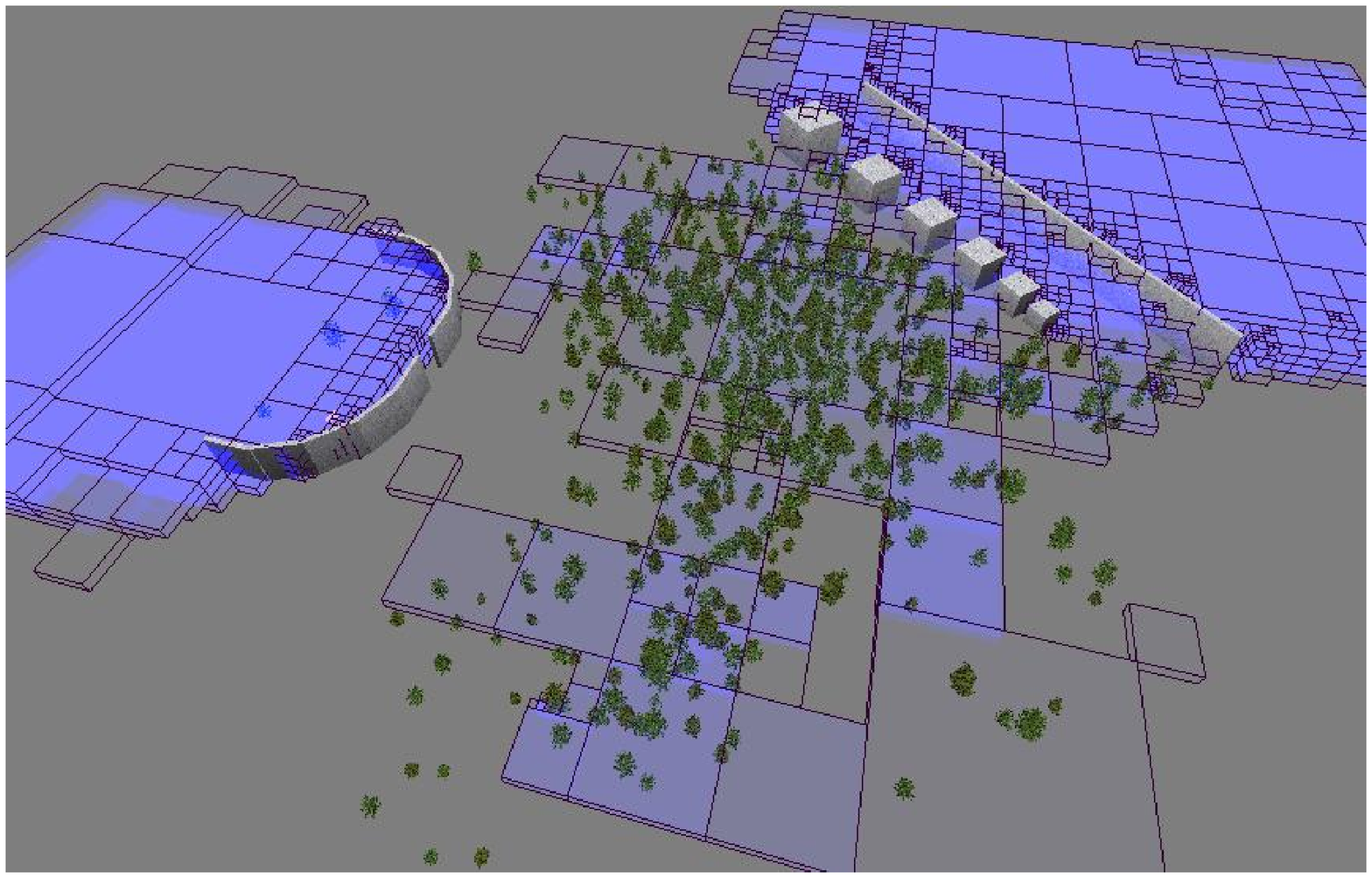}
       \caption{Area (in blue) where on the test system occlusion culling pays of, other areas are not displayed; resolution $256\times256\times8$; the scene is organized in an octree with depth 5.}
       \label{fig:runtimeDiff}
\end{minipage}%
\hspace{2mm}
\begin{minipage}[t]{0.5\textwidth}
      \centering
		   \includegraphics[width=0.85\textwidth]{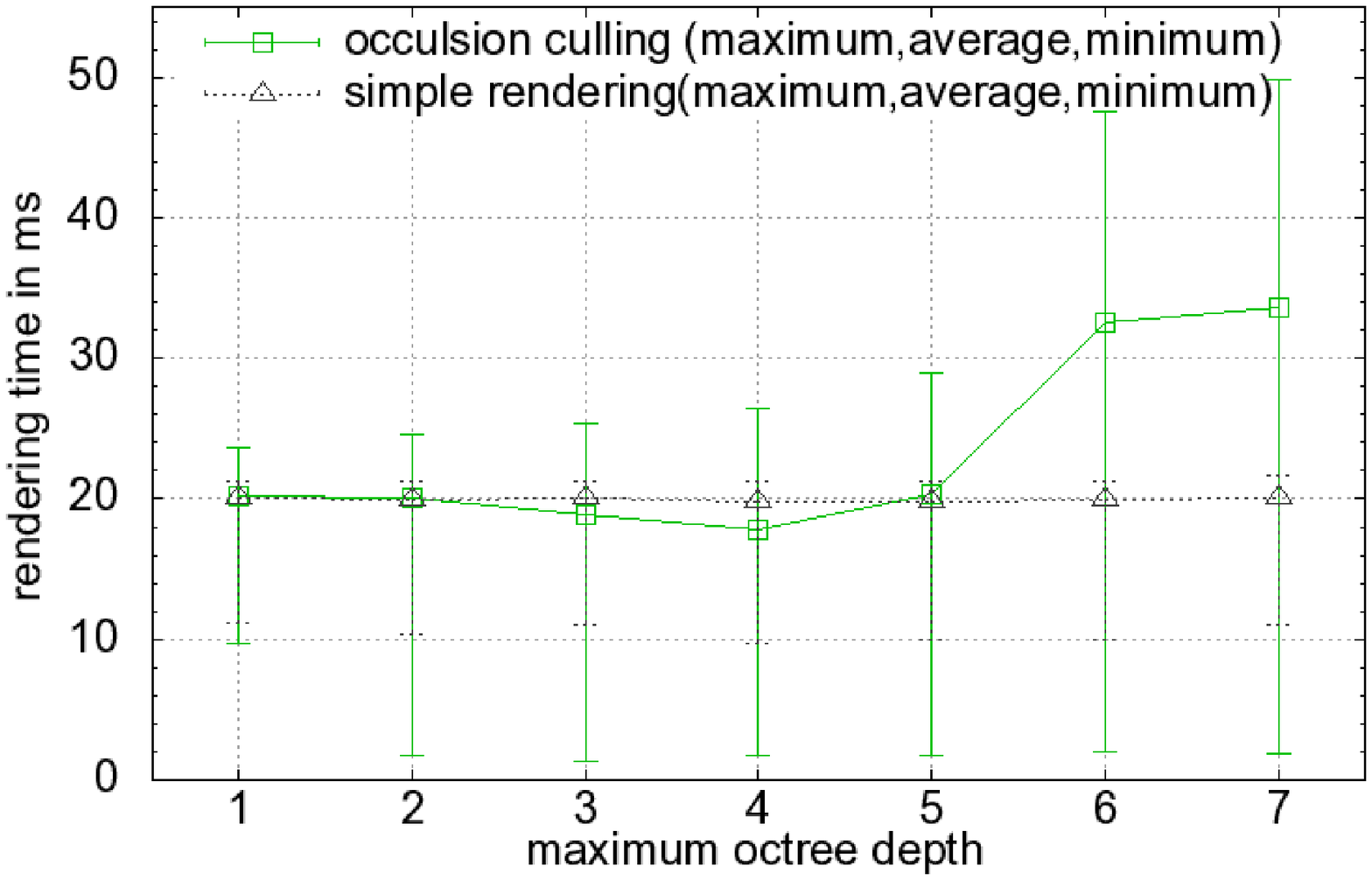}
       \caption{Average running times according to different octree depths.}
       \label{fig:optOctreedepth}
\end{minipage}%
\end{figure}

The question arises, whether the amount of culled objects is sufficient in relation to the computational overhead, introduced by the tests to  increase the overall frame rate during rendering. Therefore, we compare the subdivisions according to the running time of the two algorithms (simple rendering and occlusion culling). The function is evaluated by measuring the rendering time\footnote{Test system: Intel(R) Core(TM)2 CPU 6600 (2x2.4 GHz), 2GB RAM, Vga: ATI Radeon HD 2600 XT with 512MB RAM} at the given position for the six directions of the surrounding cube and then taking the maximum of these values (this approach emerged to identify the most relevant value in our experiments).
The difference subdivision of these two subdivisions (see Figure \ref{fig:runtimeDiff}) shows the areas, where the occlusion culling algorithm outperforms the simple rendering in blue. Inside these areas, the number of occluded (and identified as such) objects is high enough to compensate the additional costs for the tests. But if the goal is to minimize the average rendering time for all positions, then in this configuration simple rendering without occlusion culling should be preferred (average rendering time for all positions: 20.3 ms with occlusion culling, 19.7 ms without).


%
%

\section{Parameter Tuning: Maxmimum Octree Depth}\label{s:parameters}
The used occlusion culling algorithm does not perform a test for every single object, but tests the visibility of all objects that are contained in an octree node simultaneously. 
This reduces the number of tests but increases the number of occluded objects, which are rendered by mistake. 
The number of objects per node can be controlled (in our implementation) by the maximum depth of the octree, which can be declared as a parameter in the preprocessing. 
High values result in a deep octree with only few objects per node and in a good identification of visible objects.
Lower values on the other hand result in more errors during the tests, but do also reduce the number of necessary tests. 

This is one example for a parameter which can be adapted to the scene and the used hardware with the help of valued subdivisions. 
We define the optimum as the value, for which the average rendering time is minimized for the area of interest (another definition could for example be, that the area where a certain rendering time is exceeded should me minimized). 
The area, for which the subdivision is calculated should correspond to the appointed application. 
If the scene is used in a walk-through application, in which the observer mainly stays near the ground, the area of the subdivision should also only be created in this area. 
For a flight simulation, the efficiency of the algorithm above the ground may result in a different solution, as only few objects are occluded from above.

In order to determine the best value for walking through the example scene, we create multiple subdivisions according to the running time of the rendering algorithm with different values and calculate the average. Figure \ref{fig:optOctreedepth} shows the average values for the different octree depths; a value of four results in the lowest average rendering time and is therefore the best choice for the given setting according to the definition.


\section{Rendering- and Culling-Time Prediction}\label{s:prediction}
Subdivisions according to the actual runtime are a very practical tool for tuning an algorithm's parameters for a specific hardware, but they give only little insight into the internal characteristics and bottlenecks of the analyzed algorithm. 
If the dependency between the efficiency of the rendering algorithm and the properties of the hardware and the scene are understandable, the hardware can be chosen to fit the needs (e.g. in industrial applications like real time simulations) or the input (the scene) can be adapted to the abilities of a special hardware (e.g. designing scenes for computer games).
In order to create a runtime prediction for generic hardware, the first step is to identify the dominant operations of the algorithm and to formulate a function that predicts the running time in dependency of these operations (which is not a trivial task!). 
Then we create subdivisions according to the number of  operations the algorithm performs. 
At this point an (hopefully negligible) error can be introduced, as the number of operations may also be dependent on the actual hardware (especially when the algorithm heavily exploits parallelism). 
The subdivisions according to the number of operations can then be combined with the actual costs of a specific system according to the function for predicting the runtime. 
This results in a new subdivision according to the predicted running time for the test system (Probably, this process has  to be repeated a few times until the model fits the algorithm.).

\begin{figure}
\begin{minipage}[t]{0.23\textwidth}
      \centering
		   \includegraphics[width=0.99\textwidth]{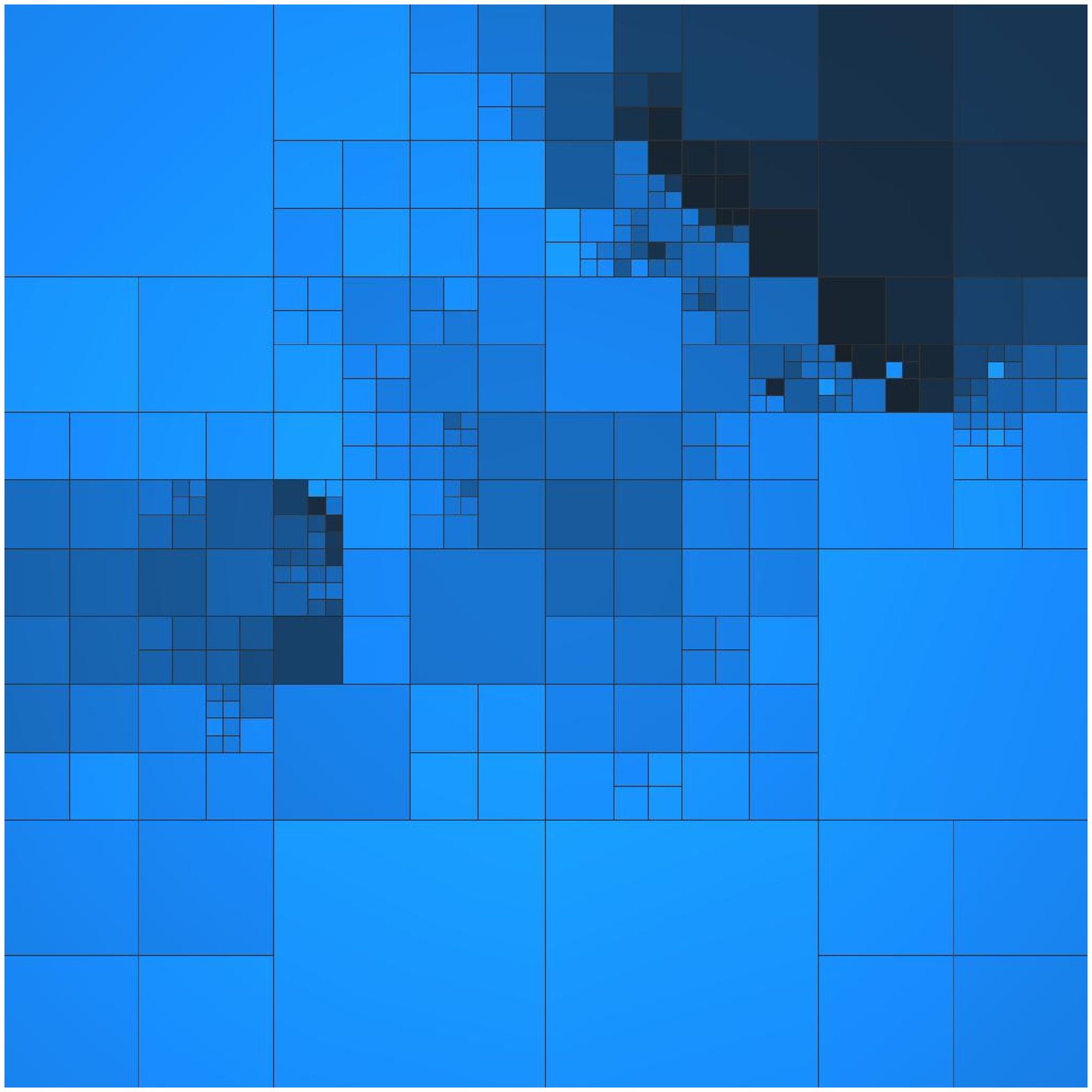}
       \caption{Subdivision according to number of occlusion tests; min (black) $20$, avg $265$, max (blue) $362$.}
       \label{fig:numTests}
\end{minipage}%
\hspace{2mm}
\begin{minipage}[t]{0.23\textwidth}
      \centering
      \includegraphics[width=0.99\textwidth]{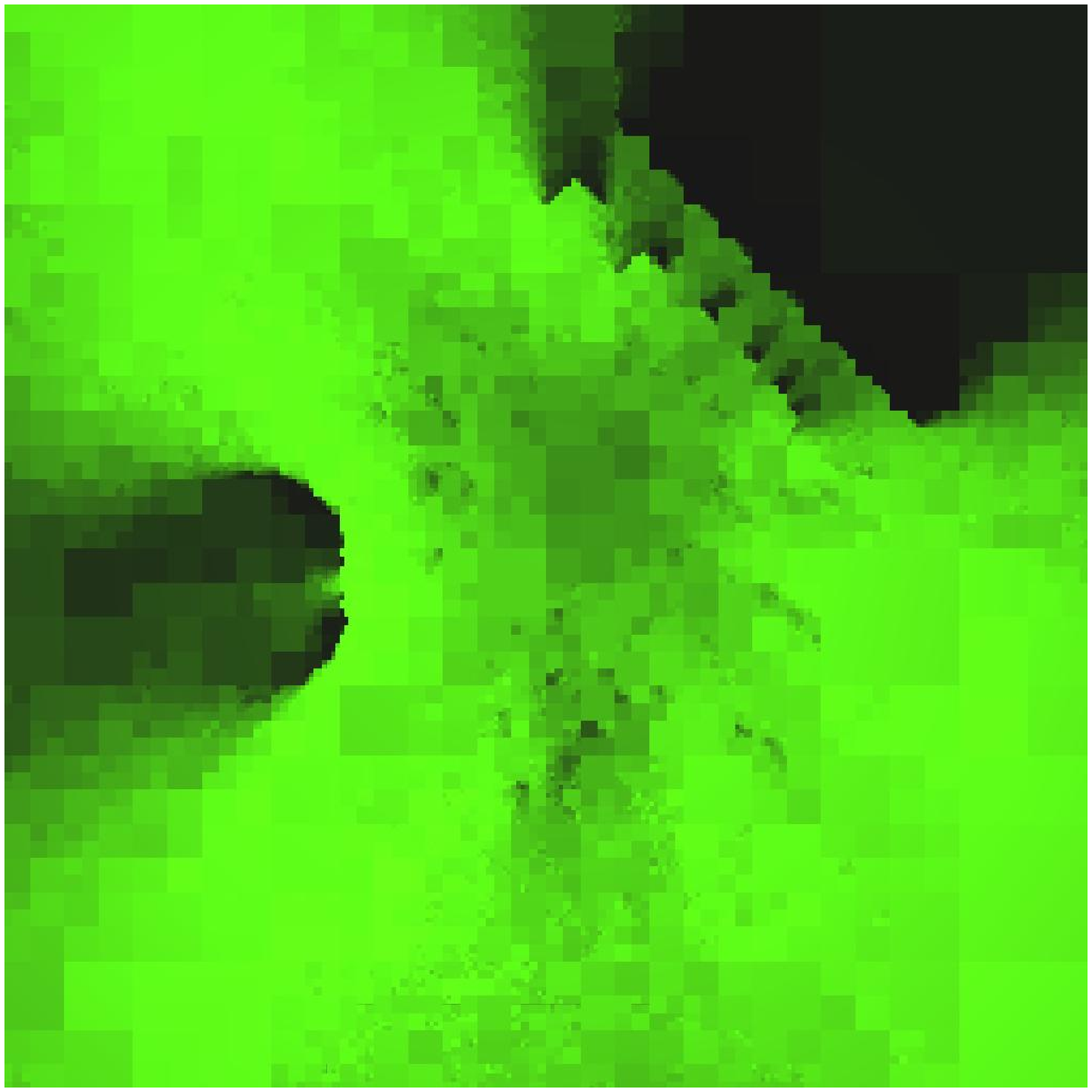}
      \caption{Subdivision according to number of rednered polygons; min (black) $84$, avg $3.4M$, max (green) $5.7M$.}
      \label{fig:numPoly}
\end{minipage}%
\hspace{2mm}
\begin{minipage}[t]{0.23\textwidth}
      \centering
      \includegraphics[width=0.99\textwidth]{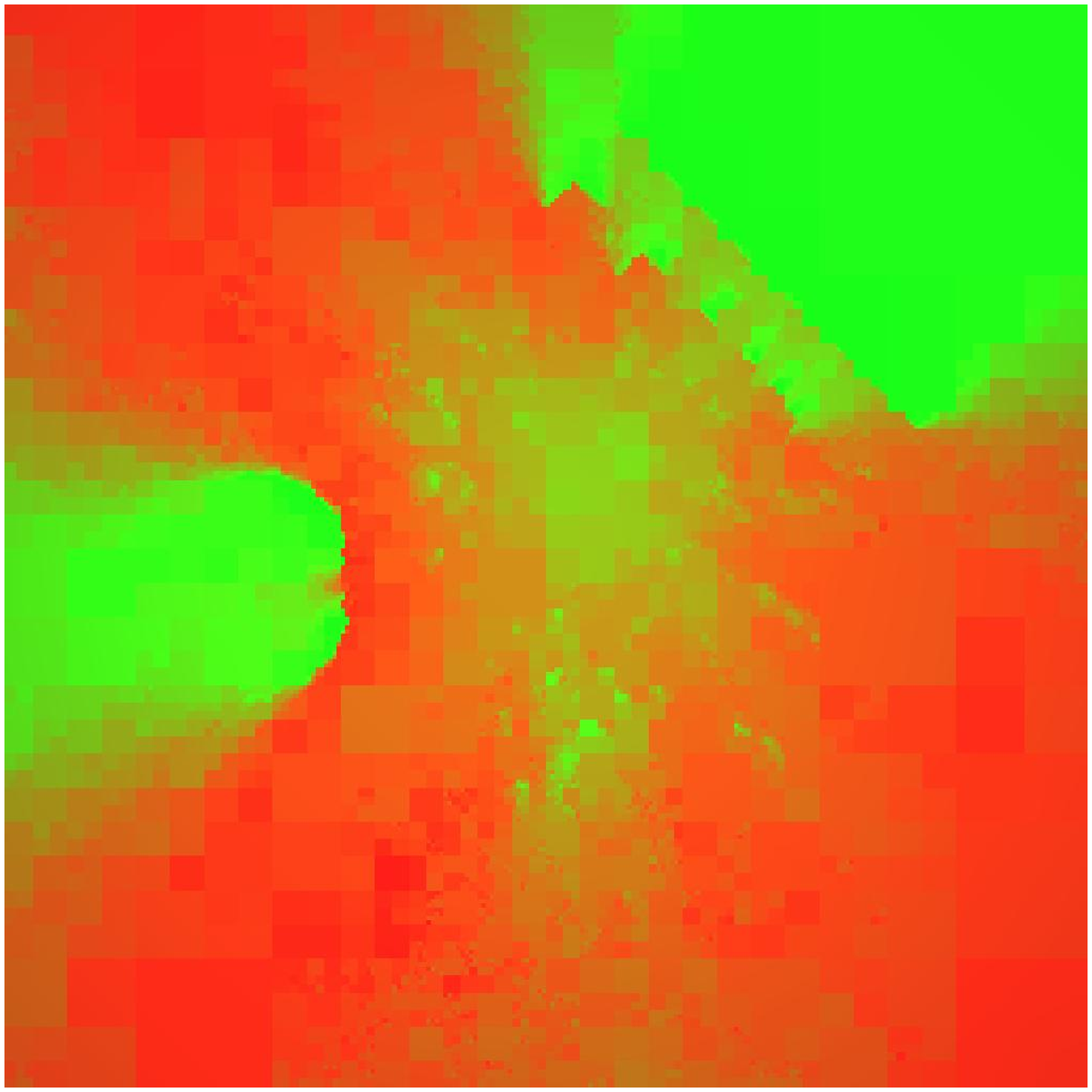}
      \caption{Subdivision according to expected rendering time; $c_{poly}$=$4*10^{-6}ms$, $c_{test}=$$0.052ms$; min (green) $1.5ms$, avg $27ms$, max (red) $40.4ms$.}
      \label{fig:expectedRendering}
\end{minipage}
\hspace{2mm}
\begin{minipage}[t]{0.23\textwidth}
      \centering
      \includegraphics[width=0.99\textwidth]{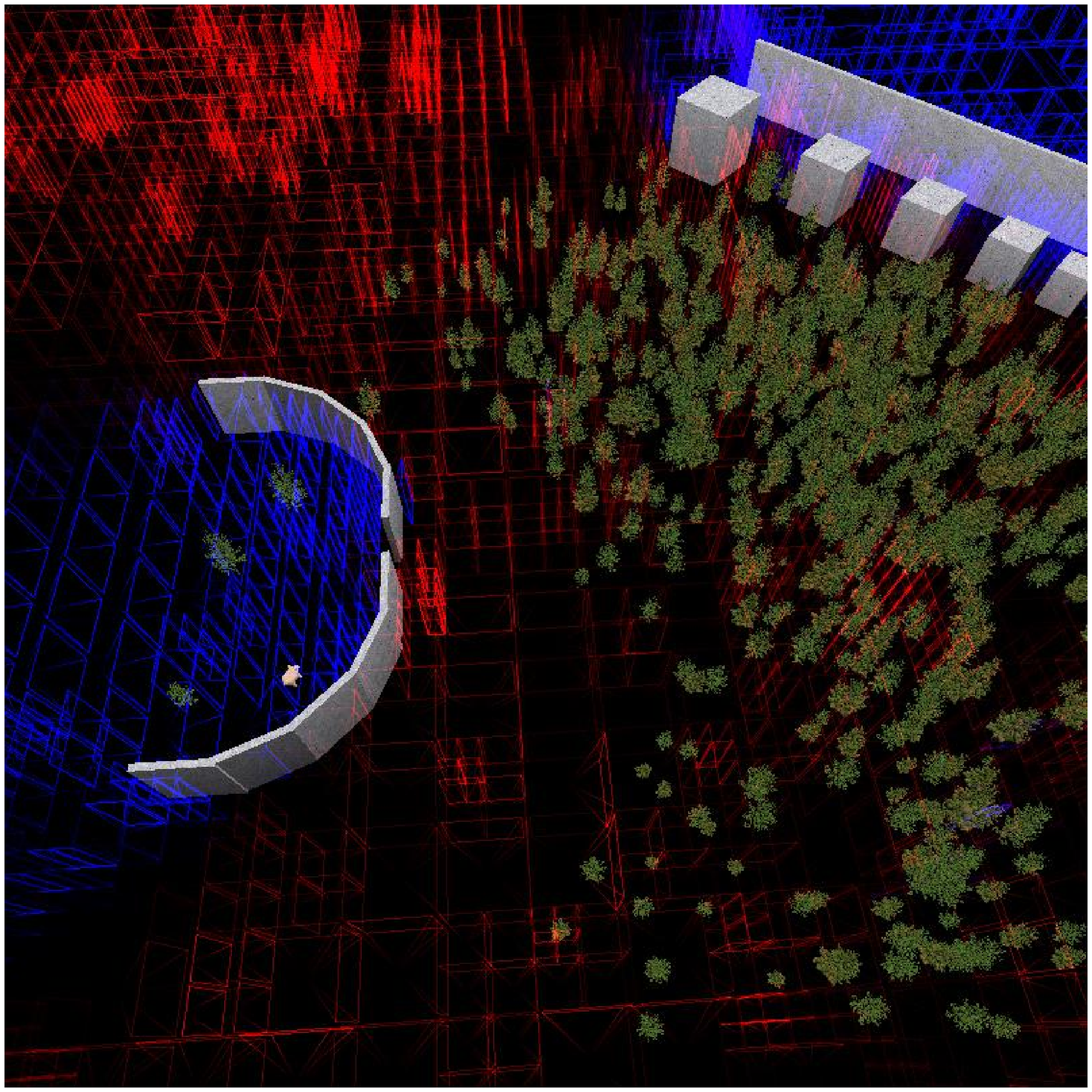}
      \caption{Part of a subdivision with viewing directions; blue: occlusion culling is faster, red: simple rendering is faster.}
      \label{fig:direction}
\end{minipage}
\end{figure}

As an example we introduce a simplified runtime prediction for the used occlusion culling algorithm: $cost_{total}(pos)=cost_{poly}*numPolygons(pos) + ost_{occlTest}*numOcclTests(pos)$. We create the according subdivisions (see Figures \ref{fig:numTests}, \ref{fig:numPoly}) for the operations and measure the average costs $cost_{poly}=4*10^{-6}ms$ and $cost_{occlTest}=0.052ms$ on the test system. 
Although we made some rough simplifications for this example, experiments showed that the results (see Figure \ref{fig:expectedRendering})  seem to be a reasonable model for the behavior of the algorithm. 


\section{Automatically and Adaptively Selecting Culling Algorithms}\label{s:selection}
We showed several possibilities to evaluate, select and adapt a rendering algorithm in the preprocessing, but it is also possible to use the additional information contained in a subdivision online during, the walk-through. 
As we have seen in Figure \ref{fig:runtimeDiff}, even in our simple example it is dependent on the position of the observer which rendering algorithm is better (in our case that means faster; but could also mean: less approximation errors, more details, etc.). 
If the behavior of multiple algorithms can be estimated by different subdivisions and the used rendering algorithms can be switched during runtime (e.g. it may be necessary that the algorithms can all work with the same data structure the scene is organized in), then we can always select the best algorithm according to the predictions. 
While only considering the actual position of the observer, this can be achieved with the presented techniques. 
But besides the position, the viewing direction of the observer has an important influence on the algorithm (although this is also true for the other applications, there it is mostly sufficient to acquire only one value per position in order to evaluate the general or average performance). 
To cover this additional information it is e.g. possible to extend the domain of the sampling by two additional dimensions for the viewing direction.
In our implementation we use another approach. 
Its main idea is to extend the dimension of the target function to six (one value for each side of the surrounding cube).
For the decision whether the current area has to be split up during the sampling process, we check the difference of each single dimension to the average value of this dimension independently. 
If the difference in one dimension is larger than the splitting threshold, the area is subdivided.   

During the walk-through, the values for the sides are interpolated according to the projected sizes of the sides on the current viewing rectangle. 
Figure \ref{fig:direction} shows a part of the visualization of the difference subdivision between the runtime subdivisions with and without occlusion culling with the viewing direction extension (for which it is quite challenging finding an acceptable visualization).  
When during the walk-through, the observer is moving through a cube and mostly sees red sides then occlusion culling is probably more efficient than simple rendering. 
When she would mainly see blue sides of the cube, the occlusion culling does not pay off and simple rendering is likely to be faster.

\section{Subdivision Quality}\label{s:practical}

When creating a subdivision, there are different parameters for adapting the result to the corresponding requirements resulting from the intended application. 
When the aim is to evaluate an algorithm in a more comprehensive way than simply using a camera path, it may not be important to have a very fine grained underlying grid and even a distinction in different viewing directions, but it may be helpful to get an intuitive visualization of the data and reliable average values. 
If a subdivision is used to select the best algorithm at runtime, the demands on the accuracy are higher.
Any decision decision should correct (or produce slight errors only) at almost all observer positions. 
Therefore a high resolution of the sampling grid, a low split value and viewing direction dependency are necessary.

\begin{figure}
\begin{minipage}[t]{0.50\textwidth}
      \centering
		   \includegraphics[width=0.99\textwidth]{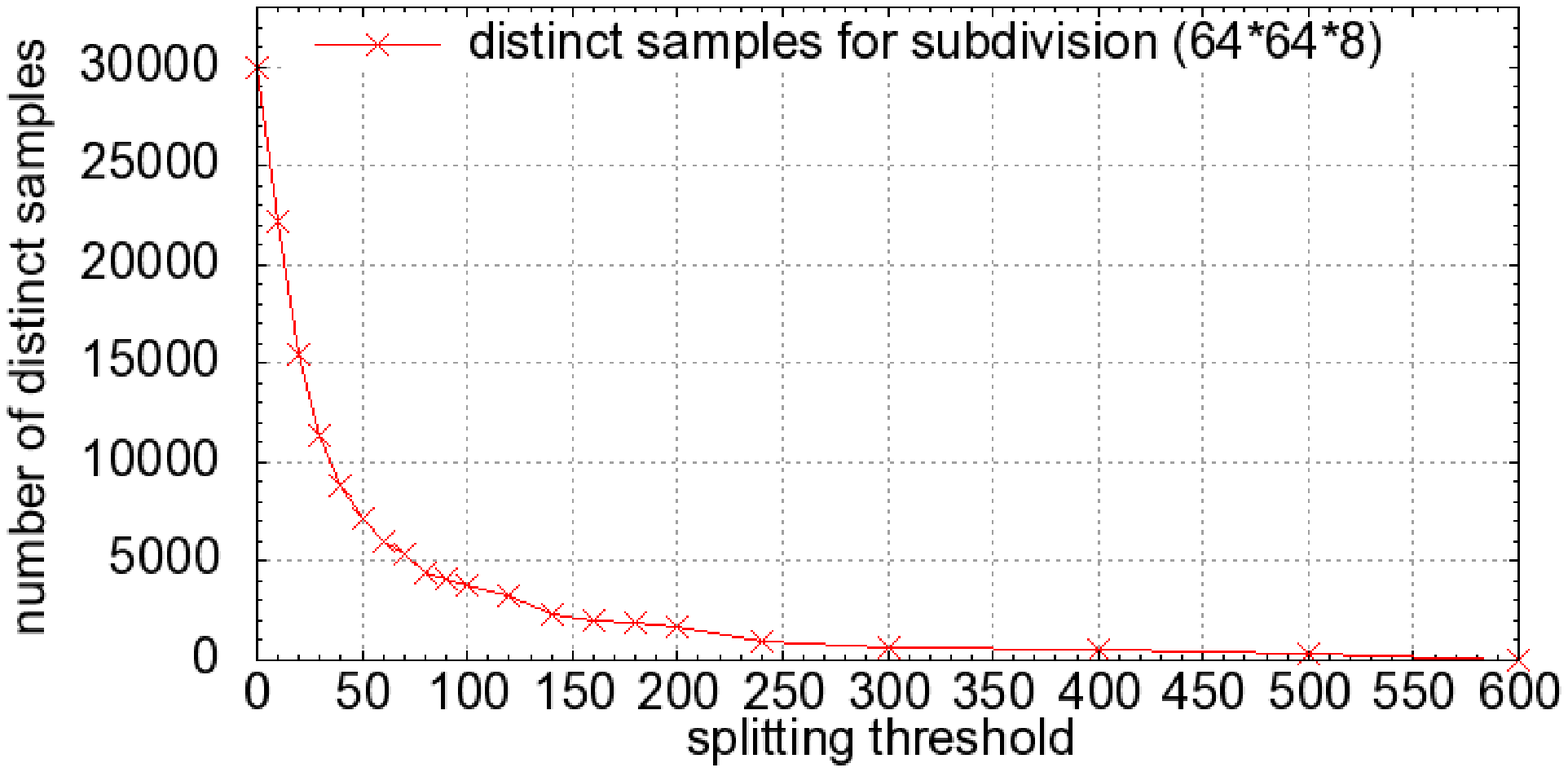}
       \caption{Number of distinct samples needed for creating subdivision according to visibility with a grid resolution of $64\times64\times8$.}
       \label{fig:numSamples}
\end{minipage}%
\hspace{2mm}
\begin{minipage}[t]{0.50\textwidth}
      \centering
      \includegraphics[width=0.99\textwidth]{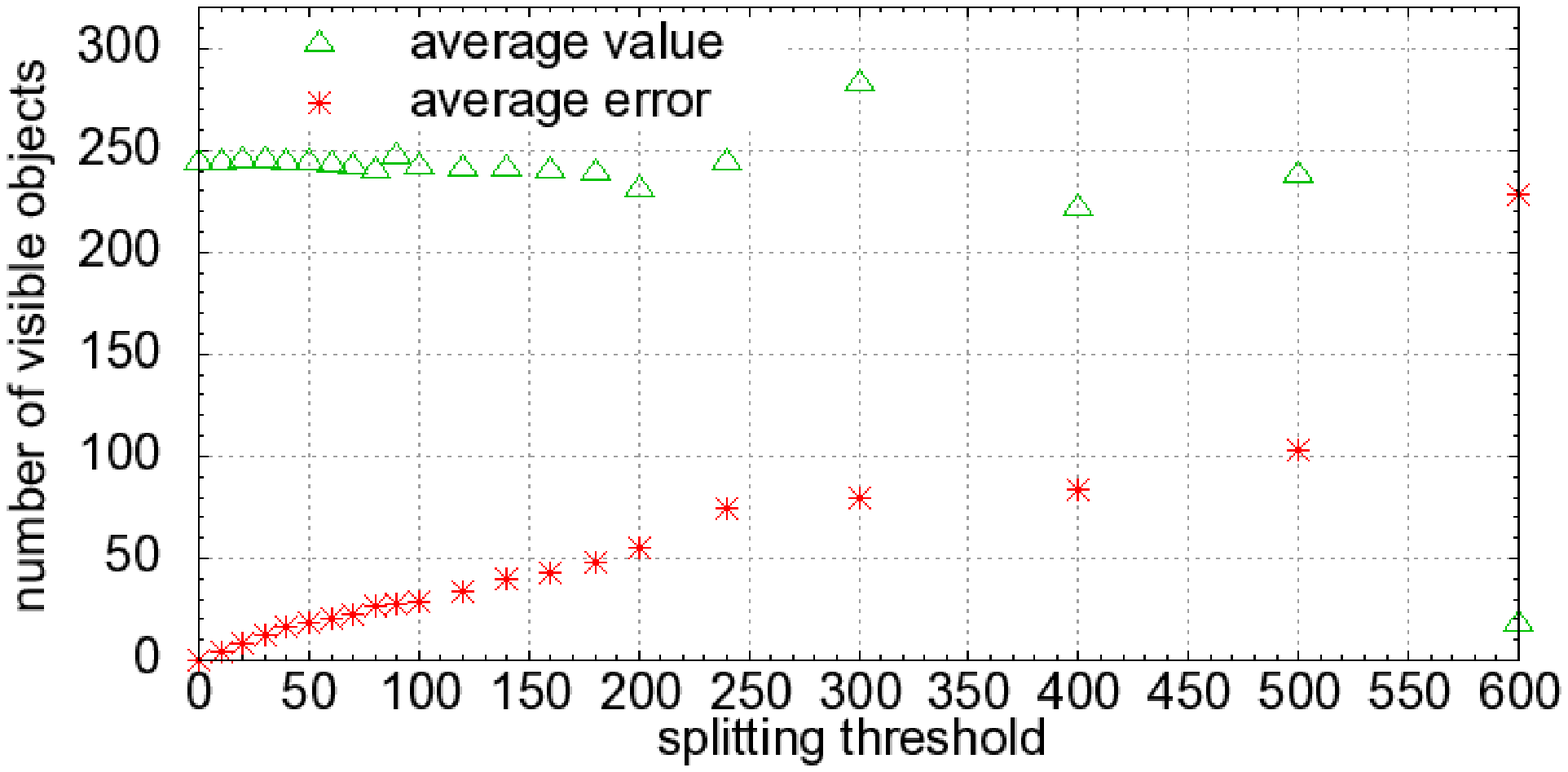}
      \caption{Average value of the subdivision and average error compared to the exact value of each grid cell in dependency to splitting threshold}
      \label{fig:error}
\end{minipage}%
\end{figure}

Figure \ref{fig:numSamples} shows the number of distinct samples needed for the calculation of a subdivision (every function value is internally cached, so that the running time is mainly determined by the number of distinct values). The time for creating this example subdivision reach from over 30 minutes (splitting threshold 10) to 20 seconds (splitting threshold 500).
Figure \ref{fig:error} shows the development of the average value and the error according to the value calculated for every grid point. 
In our experiments even a high splitting threshold leads to a good average value. 
If a low average error is important at every position, a lower threshold has to be chosen (procedure: first choose high splitting threshold, lower iteratively until result is satisfying).

\section{Conclusion and Perspectives}
We have shown the benefit of our approach in order to
automatically determine whether and when
a specific hardware-based occlusion culling algorithm
yields a net benefit over a brute-force renderer.
In the future we will extend this method from this on/off-problem
towards a finer tuning: passing to the culling algorithm
a parameter controlling how careful it is to filter (i.e. how
much computational effort to spend on finding) occluded objects
in order to yield the best net performance.
Algorithm~\ref{a:Claudius} approximates (and succeeds with high
probability on) an unknown Lipschitz-continuous function $f$ by
replacing it with piecewise constant $g$. Inspired by the works
\cite{Cooper,Beliakov}, it seems promising
to generalize our approach and use a piecewise linear $g$.



\end{document}